\documentclass[12pt,letter,final]{iopart}

\usepackage{iopams}  
\usepackage{graphicx}
\usepackage{epstopdf}
\usepackage[breaklinks=true,colorlinks=true,linkcolor=blue,urlcolor=blue,citecolor=blue]{hyperref}
\usepackage[numbers,sort&compress]{natbib}

\begin{document}

\title[Heating and ion transport in a Y-junction surface-electrode trap]{Heating rates and ion motion control in a Y-junction surface-electrode trap}

\author{G Shu$^{1}$, G Vittorini\footnote{Current Address: Joint Quantum Institute, University of Maryland Department of Physics and
National Institute of Standards and Technology, College Park, MD 20742, USA}$^{1}$, C Volin$^{2}$, A Buikema\footnote{Current Address: Department  of Physics, Massachusetts Institute of Technology, Cambridge MA 02139, USA}$^{3}$, C S Nichols\footnote{ Current Address: Department of Electrical Engineering, Princeton University, Princeton NJ 08544, USA}$^1$, D Stick$^{4}$ and Kenneth R. Brown$^{1}$}
\address{$^1$ Schools of Chemistry and Biochemistry, Computational Science and Engineering, and Physics, Georgia Institute of Technology, Atlanta GA 30332, USA}
\address{$^2$ Georgia Tech Research Institute, Atlanta GA 30332, USA}
\address{$^3$ Department of Physics, Haverford College, Haverford PA 19041, USA}
\address{$^4$ Sandia National Laboratories, Albuquerque, NM 87185, USA} 
\ead{shugang@gatech.edu}

\begin{abstract}
We measure ion heating following transport throughout a Y-junction surface-electrode ion trap. By carefully selecting the trap voltage update rate during adiabatic transport along a trap arm, we observe minimal heating relative to the anomalous heating background. Transport through the junction results in an induced heating between 37 and 150 quanta in the axial direction per traverse. To reliably measure heating in this range, we compare the experimental sideband envelope, including up to fourth-order sidebands, to a theoretical model. The sideband envelope method allows us to cover the intermediate heating range inaccessible to the first-order sideband and Doppler recooling methods. We conclude that quantum information processing in this ion trap will likely require sympathetic cooling in order to support high fidelity gates after junction transport. 
\end{abstract}


\vspace{2pc}

\section{Introduction}       
\label{sec:intro}

 One promising method for building scalable ion trap quantum computers is the ion charge coupled device architecture \cite{Kielpinski-2002}. This architecture requires an array of electrodes capable of trapping ions at arbitrary locations, shuttling ions through two-dimensional junctions, and merging and splitting ion chains. Surface-electrode ion traps ~\cite{Chiaverini2005,Seidelin2006,Chuang-PlanarTrap,Leibrandt2009,Britton2009, Stick2010a, Allcock2011, Moehring2011, Wright2013, CharlesDoret2012, Haffner2011} are well suited to all of these operations and precise micro-fabrication allows for complicated and scalable trap electrode geometries. These traps support higher secular frequencies than macroscopic traps and can be outfitted with additional features such as integrated near-side or on-chip optics \cite{Streed2009, WinelandFiber-2010, TrueMerrill2011, SandiaFiber2011} and microwave control lines \cite{Allcock2013, Warring2013, Shappert2013}. A primary disadvantage of surface traps is the larger anomalous heating rate, though this heating can be reduced through cryogenic cooling \cite{antohi2009, Vittorini2013, Poitzsch1996} or in-situ surface cleaning \cite{Hite2012, Daniilidis2011}.

We have investigated one such trap whose defining characteristic is a central Y-junction. The trap loading and transport characteristics have been previously reported in Ref.~\cite{Moehring2011}; however, ion heating data was not. In this paper, we demonstrate ground state cooling of an ion in the trap, determine the stationary heating rate due to anomalous heating, and measure heating following transport in linear and junction regions of the trap. We focus on axial heating since axial modes are commonly used for two-qubit gates~\cite{Leibfried2003, Haljan2005, Home2006, Toyoda2010, Schindler2013}.

The paper is organized as follows. In Section~\ref{sec:device}, we describe our surface-electrode ion trapping system. In Section~\ref{sec:basics}, we describe the basic trap operations. In Section~\ref{sec:heating}, we present heating rate measurements for a stationary ion, ion motion in the linear region, and ion motion through the junction. In Section~\ref{sec:conclusion}, we conclude with a discussion of trap robustness and potential future directions.

\section{Experimental Apparatus}
\label{sec:device}

\subsection{Trap, Vacuum Chamber, and Trap Electronics}
\begin{figure} [htb!]
\begin{center}
\includegraphics[width=\linewidth]{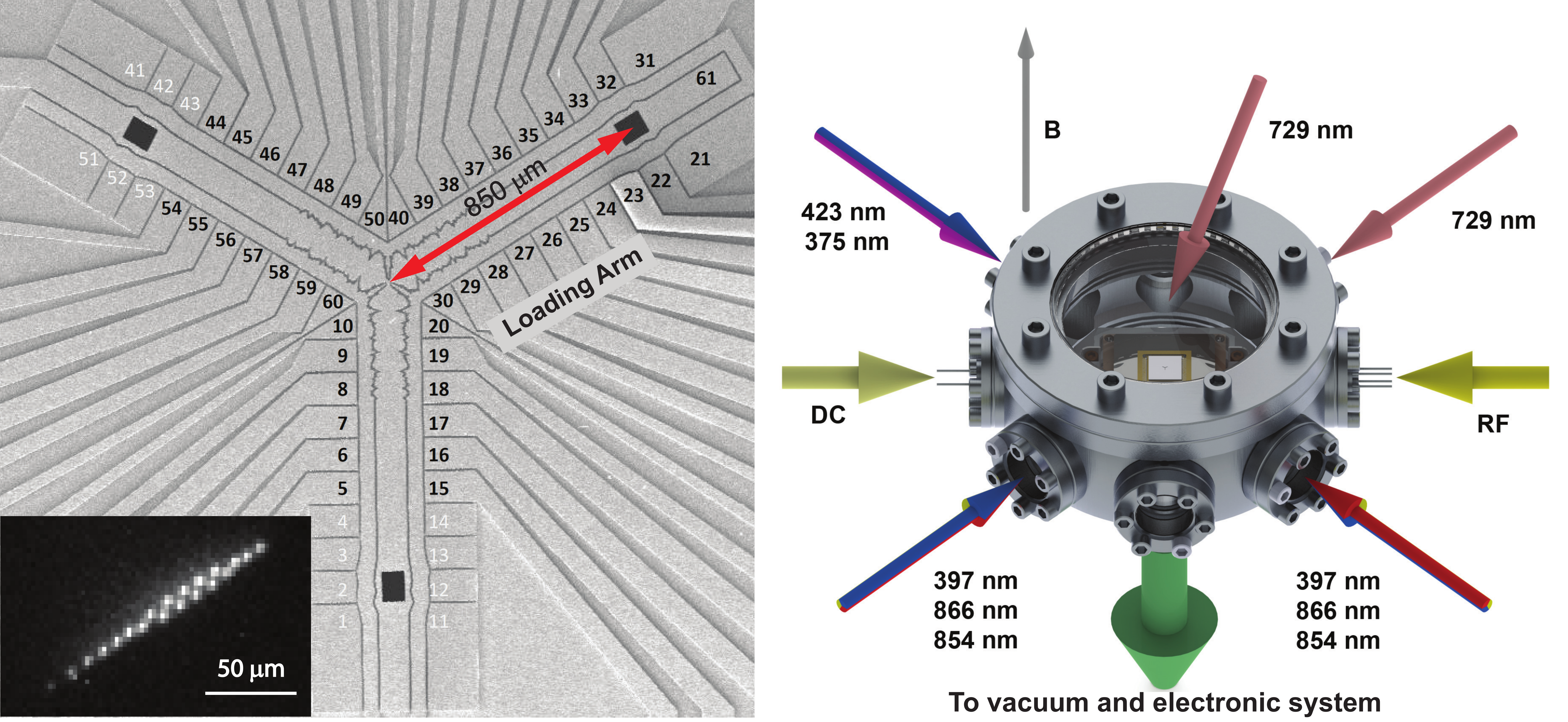}
\caption{Left: The Sandia Y-junction surface-electrode trap: $47$ of the $61$ DC electrodes~(black labels) are connected to the DACs, and the reminder (white labels) are grounded. Right: Optical, electrical, and vacuum access. The trap and the chamber are shown in the same orientation. Inset: An ion crystal held above the loading aperture.}
\label{fig.trap}
\end{center}
\end{figure}

The aluminum coated Y-junction surface-electrode trap was fabricated by Sandia National Laboratories using a foundry process as described in Ref.~\cite{Moehring2011} (type YH). As shown in Fig.~\ref{fig.trap}, the trap has $61$ DC electrodes, one radio frequency (RF) electrode, and three $70$~$\mu$m $\times$ 86~$\mu$m square through apertures for loading. Out of the $61$, DC electrodes $47$ are independently controlled and the remaining $14$ are grounded. The trap is mounted on a CPGA-100 package and gold wire bonds are used to connect the electrode traces to the package. 

The chip package interfaces with a customized polyether ether ketone~(PEEK) CPGA-100 socket inside a $4.5"$ spherical octagon (Kimball Physics MCF450-SphOct-E2A8). Among the $100$ pins, $96$ are connected to four Sub-D $25$ pin UHV compatible feedthroughs via one-foot long Kapton wires. The RF trapping signal is fed into the chamber by a UHV compatible power feedthrough (Kurt J. Lesker EFT0142052). A $40$~L/s diode ion pump (VacIon Plus 40) maintains the pressure below $10^{-10}$~torr. In order to shield the ions from potential charging of the top viewport, a thin, grounded stainless steel mesh is mounted $2$~mm above the trap surface. There are six $0.75"$ side viewports for laser access, though the location of the mesh shield mount restricts laser access.

The $45$~MHz trapping RF signal is generated by either a commercial function generator (BK Precision 4087) or by one channel of an Analog Devices AD9958 Direct Digital Synthesizer (DDS) board. This signal is then increased by 35 dB using a Mini-Circuits TIA-1000-1R8-2 instrument amplifier to $\le1.5$~W before the helical resonator. Twelve National Instruments (NI) PXI 6713 digital-to-analog converter~(DAC) cards are used to generate DC trapping voltages. They have an output voltage range between $\pm10$~V. When synchronized by a single clock signal, the $96$ output channels can be updated simultaneously at a maximum rate of $750$~kHz. In order to generate higher axial secular frequencies and have increased flexibility for shuttling and merging waveforms, a Texas Instruments OPA445 high voltage operational amplifier circuit with a gain of 3$\times$ is added to each analog channel. All DC lines are filtered on the air side of the vacuum feedthrough by a single pole low-pass filter with a cut-off frequency of $60$~kHz. Both the static and dynamic noise of the final output channels are measured with an HP8591E spectrum analyzer to be below $-70$~dBm between $100$~kHz and $2$~MHz.

\subsection{Ion loading and laser cooling}
Ions are loaded by photoionization from a  neutral atomic vapor, which is generated by resistively heating a stainless steel tube containing calcium metal mounted below the chip holder. At a current of $2$~A, neutral fluorescence can be observed by driving the Ca $4s^2$~$^1S_0\rightarrow4s4p$~$^1P_1$ transition with 25~$\mu$W of 422.6~nm light focused to a waist of 50~$\mu$m. The 422.6 nm light is generated from a BBO crystal in a resonant cavity (Toptica SHG110) pumped by an $846$~nm external cavity diode laser (ECDL, Toptica DL100 for all instances in this report). The excited calcium atoms can be efficiently ionized by a free-running $375$~nm diode laser. Different isotopes can be selectively loaded by shifting the $422.6$~nm laser frequency ~\cite{LucasPRA2004, TanakaAPB2005, Goeders2013}. In this paper, we use $^{40}$Ca$^{+}$, unless otherwise stated. 

The trapped ions are Doppler cooled by a combination of $397$~nm and $866$~nm laser light.  The $397$~nm light driving the Ca$^+$~4s~$^2S_{1/2}\rightarrow$4p~$^2P_{1/2}$ cooling transition is derived from a frequency-doubled $794$~nm ECDL laser. A $20$~$\mu$W beam of $397$~nm laser light is combined  with $300$~$\mu$W of  $866$~nm light (Ca$^+$~3d~$^2D_{3/2}\rightarrow$4p~$^2P_{1/2}$ re-pump transition) and the beams are focused to a waist of $50$~$\mu$m and $80$~$\mu$m, respectively. The cooling beams are delivered to the trap along two perpendicular directions~(Fig.~\ref{fig.trap}). The geometry of the chamber viewports and mesh shield supports are incompatible with aligning the laser beams parallel or perpendicular to the trap loading arm. The minimum attainable angle between a laser beam and the loading arm axis is $10^\circ$.

In order to perform sideband cooling and resolved sideband heating measurements, we drive the Ca$^+$~4s~$^2S_{1/2}\rightarrow$3d~$^2D_{5/2}$ quadrupole transition with a slave laser diode injection-locked to a master 729~nm ECDL with $\le$1~kHz linewidth~\cite{Goeders2013}. The frequency drift of the laser relative to the atomic transition was measured to be less than $\pm1$~kHz over $4$~hours. At the trap site, $14$~mW of the $729$~nm light is focused to a $30$~$\mu$m waist.  The Zeeman levels are split with a magnetic field of $5$~ gauss perpendicular to the trap surface. This field is generated by running a 0.68~A current through a single coil of wire above the chamber's top viewport. The magnetic field drift shifts the Zeeman lines by $\pm 5$~kHz over $4$~hours.  

\begin{figure} [htb!]
\begin{center}
\includegraphics[width=\linewidth]{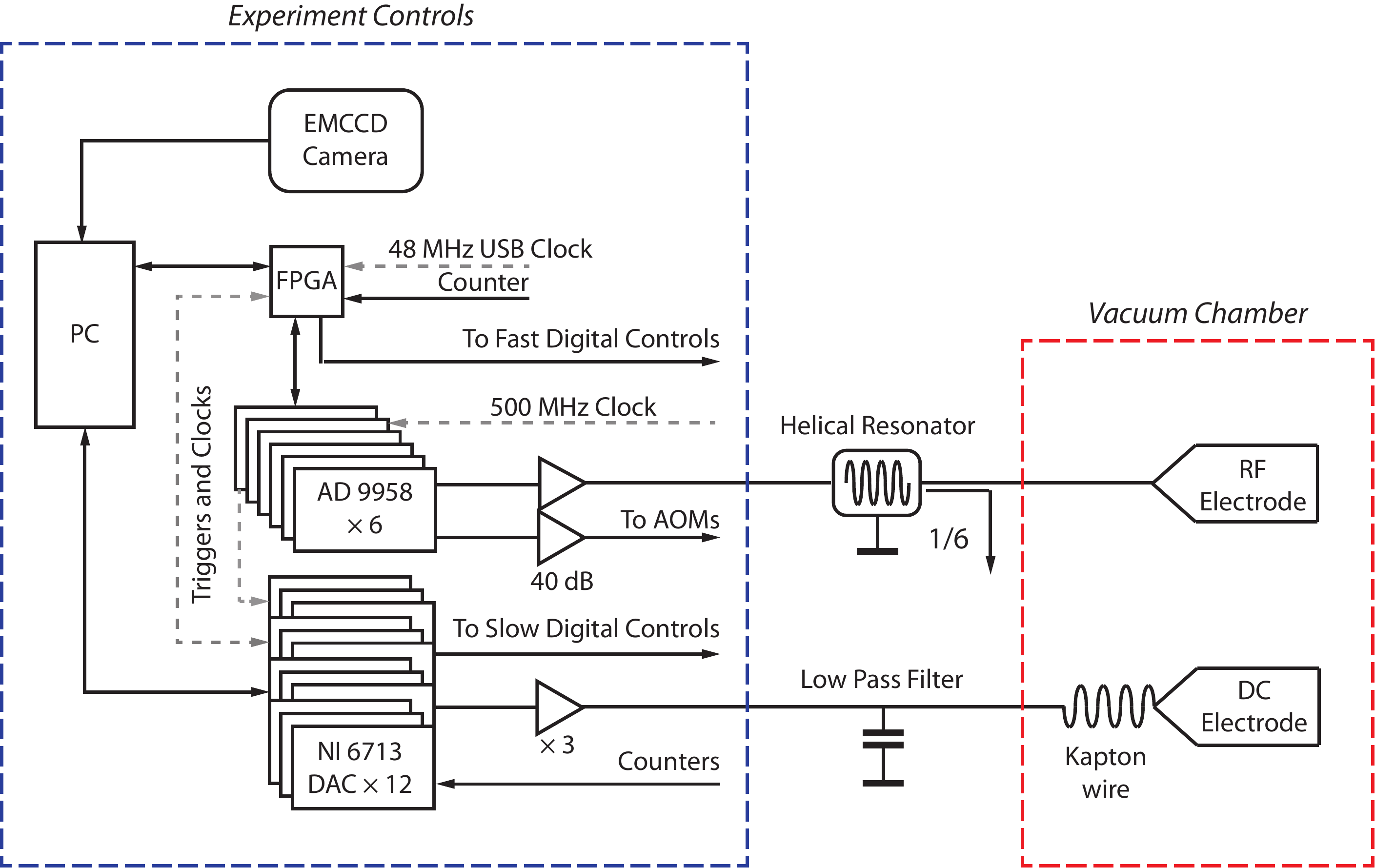}
\caption{Experimental control system: DC voltages are generated by twelve NI PXI 6713 DAC cards. An amplifier circuit~($G$ = 3) increases the voltage prior to filtering by a single pole low-pass filter~($f_c$ = 60 kHz). Six AD9958 DDS boards generate $\le$~250~MHz signals for all necessary AOMs. The 45~MHz RF trapping signal is also generated by a DDS, amplified by a 40~dB RF amplifier, and filtered by a Q$\sim$100 helical resonator. Approximately $1/6$ of the RF amplitude following the resonator is sampled to monitor the applied RF trap voltage. Custom control software manages all hardware subsystems and digital triggers generated by an FPGA are used to coordinate the various subsystems.}
\label{fig.dacsetup}
\end{center}
\end{figure}

Time-critical experimental sequences, including sideband cooling and state detection, are performed by a Xilinx Spartan 3 field-programmable gate array (FPGA) module controlling three Analog Device AD9958 DDS boards. A microprocessor core on the FPGA arranges the laser modulation signals and synchronizes the trap DC potentials via digital triggers or shared clock signals~(Fig.~\ref{fig.dacsetup}).          

\section{Trap operation and compensation}
\label{sec:basics}

Typical trap secular frequencies are 1-1.8 MHz axially and 4-6 MHz radially; these frequencies are measured via sideband interrogation and verified by driven ion oscillation. The dark ion lifetime in this trap is 30 seconds at a pressure of $10^{-10}$~torr, which is typical for surface-electrode traps.

Ions in microscale traps are generally more susceptible to anomalous heating due to the reduced distance between ions and electrode surfaces~\cite{Turchette2000a, Deslauriers2006}. This effect is enhanced around the trap loading slot due to the atomic flux from the oven coating the aperture edges and subsequent charging due to incident laser beams. Initial loading attempts in this Y-junction trap resulted in direct loading of large $^{40}$Ca$^{+}$ crystals~(Fig.~\ref{fig.trap} inset), indicating no significant stray field. Following two years of continuous operation, crystal size is currently limited to $\le$~4 ions at the same location with the same applied fields. Ion displacement measurements reveal a potential `bump' near the lower edge of the loading hole, caused by both RF ripple and a DC stray field, which axially pushes ions away from the loading slot in both directions.

Initially, ions could be transported throughout the trap and through the junction repeatedly~($\ge$5000 times) with only Doppler cooling at a single trap location. After more than two years of operation, we are still able to shuttle a single ion through the junction continuously for $\ge10^5$ times. However, to ensure reliable linear shuttling and merging of ions from separate trapping potentials, we now apply a radial compensation electric field on the order of $\sim500$~V/m along the loading arm.

We compensate micromotion due to stray electric fields via three methods. For radial motion parallel to the trap surface, we measure micromotion using the RF photon correlation method~\cite{Wineland-MMC} and by detecting resolved micromotion sidebands on the 397~nm transition~\cite{CharlesDoret2012, Vittorini2013}. For motion perpendicular to the trap surface, we measure the micromotion sidebands of the 4s~$^2S_{1/2}\rightarrow$3d~$^2D_{5/2}$ transition~\cite{Vittorini2013}. To account for the laser's spatial mode distribution, both first- and second-order sidebands are measured and compared to locate the common zero point. The axial DC stray field is compensated by minimizing ion displacement at different axial secular frequencies. Using these methods, we eliminate stray linear electric fields to within $\pm$10~V/m along all three principal axes at any single location along the loading arm between (E27, E37) and (E23, E33).  

All three compensation schemes require generating specific waveform basis functions to create bias electric fields along arbitrary directions. Instead of directly iterating the potential of individual electrodes, we use these basis waveforms to generate fields largely along one of the three principal axes with minimal projection along the other two directions. With reduced correlation between different field directions, only a few iterations~(normally $\le3$) are enough to completely compensate the stray field in all three directions. Compensation at the junction is difficult due to static ion instability within the region. However, this does not prevent shuttling through the junction.

We are unable to merge ions using the waveforms of Ref.~\cite{Moehring2011} due to uncompensated electric fields. After compensation, we can successfully merge ions along the only fully connected arm of the Y-junction trap. There are 9 pairs of DC electrodes along the $850$~$\mu$m segment from the center of the loading hole to the junction. To avoid the large stray electric fields and gradients around the loading hole and complicated geometry within the junction, we merge separately trapped ions in the middle of the arm at location (E26, E36). Limited optical access prevents cooling the whole arm with a single axial beam. Instead, we utilize two beams: one is primarily aligned perpendicular to the arm axis to address the loading zone, and the other, oriented at $\sim$10$^{\circ}$ from the arm axis, addresses the merging/storage area. Chains of up to 4 ions are constructed by individually loading and merging ions. We have also generated multi-isotope ($^{40}$Ca$^+$ and $^{44}$Ca$^{+}$) chains using isotopically selective loading combined with merging operations.

\section{Heating during motion operations}
\label{sec:heating}
For the convenience of discussion, we divide ion heating into stationary heating and dynamic heating. Stationary heating consists of any heating of the ion while it is held at a single location. This includes known heating caused by noise on DC voltage sources, coupling of RF signals to the DC electrodes, anomalous heating due to surface effects including patch potentials, and Johnson noise. Dynamic heating occurs during transport and is in addition to stationary heating. Possible sources include additional electronic noise during DC voltage updates and parametric heating due to distortion of the DC waveform during transport.  

We use multiple methods to measure the ion motional energy: first-order sideband comparison~\cite{Diedrich1989},  motional decoherence of Rabi oscillations~\cite{Roos2000}, Doppler recooling~\cite{Epstein2007}, and the sideband envelope method, which is a fit to the peak amplitudes of multiple sideband orders. For all methods, we assume the ion is in a thermal state. For cold  ($\overline{n}<10$)  systems, the first-order sideband comparison technique is the standard measurement method. The motional decoherence of carrier Rabi oscillations is appropriate for slightly warmer systems ($10\le\overline{n}\le40$), while for hot systems ($\overline{n}\gtrsim10^3$), the Doppler recooling method is commonly used. For the last two methods, a one-dimensional model is assumed. This assumption is justified for the motional decoherence method because the radial secular frequencies are significantly higher than the axial frequency and because probe beams are only $\sim$10$^{\circ}$ off the trap axis.  The sideband envelope technique measures the amplitudes of multiple sidebands and compares the envelope to a model that assumes constant interaction time and laser intensity. Details of the model can be found in \ref{Appendix}. In the limit of incoherent excitation, one expects the sideband envelope to be described by a Gaussian whose variance is proportional to temperature \cite{Naegerl2000}; in the unresolved sideband limit this is equivalent to a line broadening due to the first-order Doppler shift. We employ a fully quantum treatment without dissipation which allows us to measure low heating based on differences in red and blue sideband heights and take into account coherent oscillations in the peaks. As shown in Section~\ref{sec:dynamic}, this method can measure $0\le\overline{n}\le500$ when sidebands up to 4th order are included.

\subsection{Stationary Heating}

Before measuring ion heating rates, we sideband cool the axial motion close to the vibrational ground state. The ion is initially Doppler cooled to $\overline{n}$ = 6-10 quanta of motional energy and then continuously sideband cooled to $\overline{n}\le0.5$ quanta. With our current $729$~nm laser intensity ($\sim10$~W/mm$^2$), we have achieved a minimum $\overline{n}$ of 0.25 quanta. This relatively high value following sideband cooling is consistent with a simple estimate based on our trap heating rate.

\begin{figure} [htb!]
\begin{center}
\includegraphics[width=\linewidth]{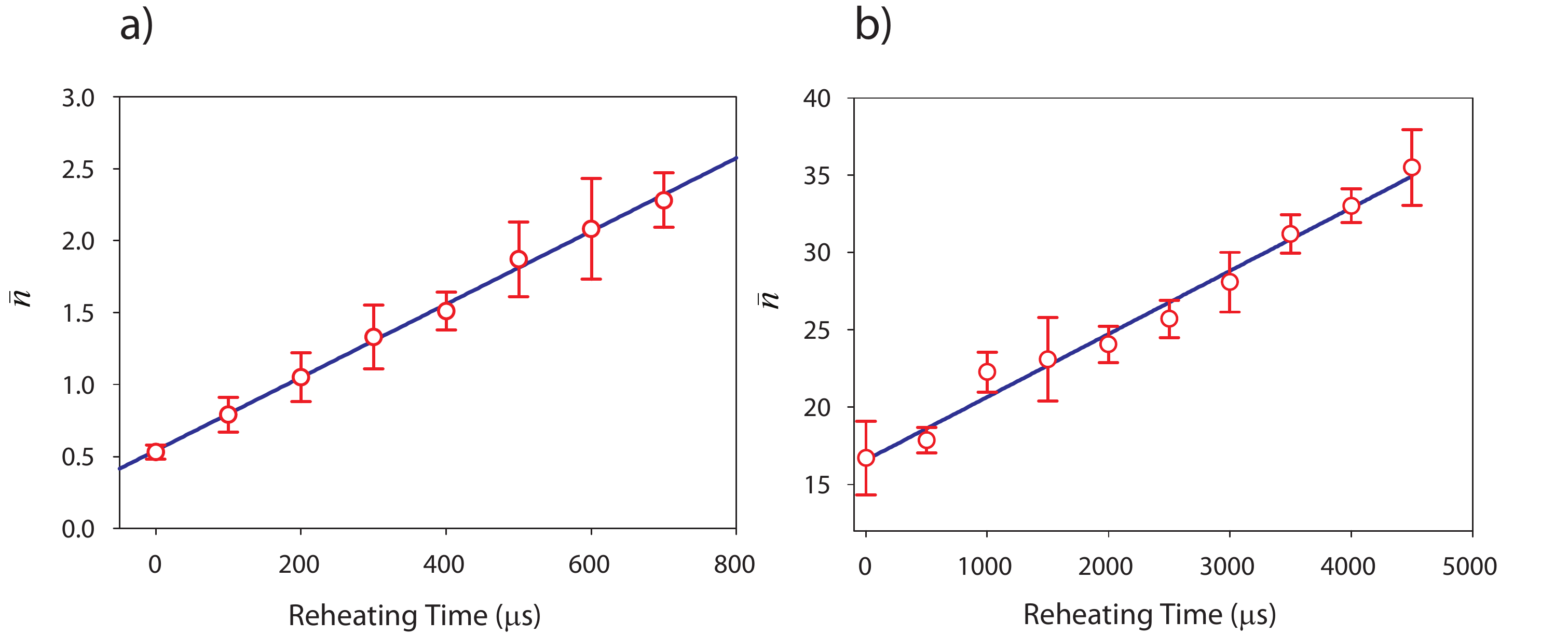}
\caption{Stationary heating rate measurements for the axial mode with a frequency of $1.738$~MHz. a) A first-order sideband comparison measurement gives a heating rate of $3.00\pm 0.06$~quanta/ms. b) Measurement of the motional decoherence of Rabi oscillations yields a heating rate of $4.0\pm 0.2$~quanta/ms.}
\label{fig.heatingrate}
\end{center}
\end{figure}

We extract a stationary heating rate of $3.00\pm0.06$~quanta/ms from the first-order sideband comparison method and $4.0\pm0.2$~quanta/ms from the motional decoherence method (Fig.~\ref{fig.heatingrate}). This agrees well given the assumptions of the motional decoherence model. At our ion-surface distance~($\sim70$~$\mu$m), this heating rate, which corresponds to a noise spectral density $\omega S\sim1.2\times10^{-3}$ V$^2$/m$^2$, is comparable to other surface-electrode traps at room temperature without surface cleaning and an order-of-magnitude worse than the best reported heating rates \cite{CharlesDoret2012}.

\subsection{Dynamic Heating}
\label{sec:dynamic}
\begin{figure} [htb!]
\begin{center}
\includegraphics[width=\linewidth]{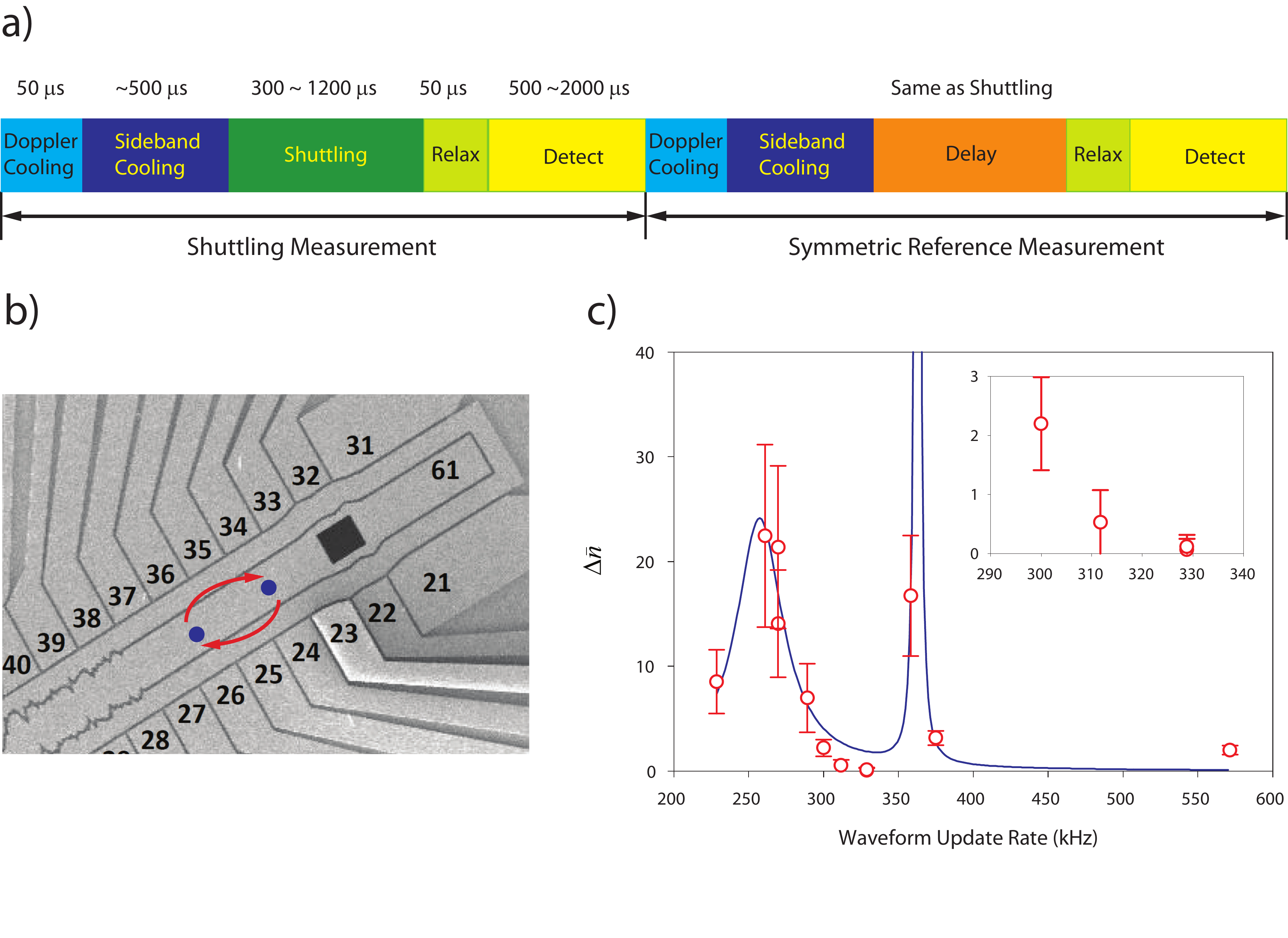}
\caption{Dynamic heating in transport: a) A single measurement cycle: the ion is sideband cooled to $\overline{n}\le0.5$ and undergoes round-trip transport. A $50$~$\mu$s pause, labeled "Relax", is inserted between the shuttling and measurement to ensure electrode voltages had reached their final, static value. A reference measurement with exactly the same elapsed time is run immediately afterwards to allow isolation of the dynamic heating. b) The ion is shuttled along the loading arm from (E26, E36) for $177.3$~$\mu$m to (E24, E34) and back. c) Ion heating during transport retrieved from first-order sideband comparison, with different waveform update frequencies, revealing two resonance peaks. A simple dual-peak Lorentzian fit shows that they are approximately $1/5$ and $1/7$ of the stationary secular frequency~(see text). The insert shows that the heating is low at update frequencies between these resonances.}
\label{fig.ShuttleHeating}
\end{center}
\end{figure}

Minimizing heating during transport is critical to maintaining motional coherence and simplifying subsequent logical operations. We measure the heating of the axial motion of a single ion after adiabatic transport and chose the middle of the loading arm for transport to avoid atypical heating due to the loading slot. Ions are shuttled from the middle of the loading arm (E26, E36) to (E24, E34) and back to (E26, E36) for an overall distance of $354.6$~$\mu$m~(Fig.~\ref{fig.ShuttleHeating}b). As the DAC update rate is limited by 60~kHz filters, the shuttling operations are classified as adiabatic. The waveform's amplitude and phase are maintained, therefore a transfer function treatment is not used (Ref.~\cite{Bowler2012}). 

In order to isolate dynamic heating effects, each measurement cycle consists of two symmetric portions (Fig.~\ref{fig.ShuttleHeating}a). One consists of ion transport followed by a heating rate measurement. The other determines the heating of an ion after sitting still for the same period of time. We look at the difference between the two extracted $\overline{n}$ to determine heating due to transport. The waveform consists of 120 steps and is swept at an update frequency of 200 - 600~kHz, corresponding to an update period of 200 - 600~$\mu$s, as the heating  can depend critically on update rate~\cite{WaltherPRL2012}. Comparing the first order sidebands, we observe a heating resonance at $257$~kHz and $363$~kHz, slightly larger than 1/7 and 1/5, respectively, of the stationary axial secular frequency (1.738 MHz) (Fig.~\ref{fig.ShuttleHeating}c). We believe the deviation from perfect subharmonic frequencies is due to the deformation of the axial trapping potential during transport. At an update frequency of 329~kHz, we see a minimal dynamic heating of $0.07^{+0.25}_{-0.07}$ quanta which is negligible when compared to the anomalous heating and a vast improvement over the 22 quanta measured near the heating resonance.

\begin{figure} [htb!]
\begin{center}
\includegraphics[width=\linewidth]{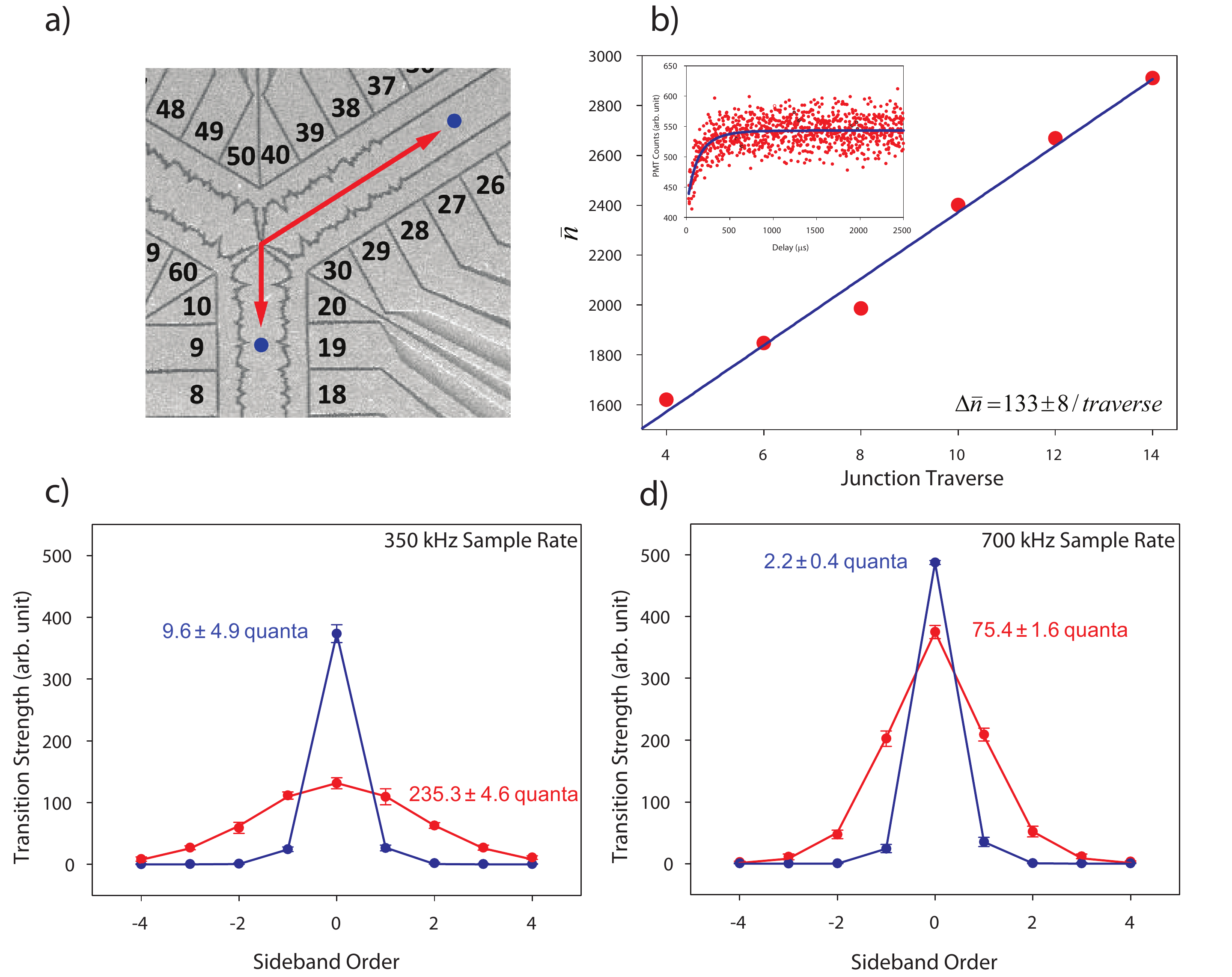}
\caption{Ion heating during junction traverse: a) The ion is shuttled through the junction and back for measurement. The overall path length is $1160$~$\mu$m. b) Doppler recooling measurement of ion heating after shuttling at a waveform update rate of 330~kHz. This method is not reliable for the intermediate heating regime, which results in an unrealistically large offset value for $\overline{n}$. A fit to multiple  junction traverses gives a heating of 266 quanta per round-trip. Inset: Recooling data after 6 junction traverses fit to the numerical model. c) At an update rate of 350 kHz, the sideband envelope method gives $\overline{n} = 9.6\pm4.9$ for the reference run (blue), and $\overline{n} = 235.3\pm4.6$ for the shuttling run (red). The dots are experimental data and the lines join the points of the envelope fit. This measurement is in agreement with the average heating rate measured by the Doppler recooling method at a similar update rate. d) A minimal dynamic heating of $73\pm2$ quanta was measured by the sideband envelope method at a waveform update rate of $700$~kHz.  The fit gives $\overline{n}= 2.2\pm0.4$ for the reference run (blue), and $\overline{n}= 75.4\pm1.6$ for the shuttling run (red).  }
\label{fig.JctHeat}
\end{center}
\end{figure}

We also measure ion heating due to transport through the junction. To traverse the junction reliably, an axial secular frequency of approximately 1 MHz is required. Static ion stability in the junction is limited by extremely high micromotion, making compensation of the junction impractical. Nevertheless, we are able to perform 20 round-trip junction traverses reliably without any laser cooling. The stationary axial secular frequency is varied from 1.46 MHz to 1 MHz during the course of the transport in order to maximize ion survival and reduce measurement noise. The waveform, calculated with the GTRI waveform package used in Ref.~\cite{Wright2013}, contains $720$ steps and shuttles the ion from (E26, E36) to (E9, E19) and back~(Fig.~\ref{fig.JctHeat}a). The shuttling operation requires between 0.96~ms to 3.6~ms for full transport, depending on the update rate. The corresponding stationary heating at (E26, E36) is between 3 and 11 quanta per trip. 

Initial heating measurements were performed at an update rate of 330 kHz.  Sideband comparison showed first order peaks of statistically equal height and the motional decoherence of carrier Rabi oscillations yielded data approaching a perfectly damped oscillator. Based on our models, this suggests an induced heating of $\Delta\overline{n}>50$ per trip. The Doppler recooling method returned a single round trip heating of $\overline{n}\sim 1200$. However, multiple trips were used to extract $\Delta \bar{n}$ and determine an average heating rate of $133\pm 8$~(Fig.~\ref{fig.JctHeat}b). This is below the limit of reliability of the Doppler recooling method and we treat the initial measurement as an unphysical offset.

In order to measure heating in this intermediate regime ($50\le\overline{n}\le500$), we employ the sideband envelope method. This method covers a wide range of $\overline{n}$ by considering both the transition ratio of symmetric sideband orders for low $\overline{n}$ and the transition ratio between different sideband orders for high $\overline{n}$. The reference run reveals low static heating, $\overline{n} = 2.2\pm0.4$, which is consistent with our single-order sideband measurement.  At a $350$ kHz update rate ~(Fig.~\ref{fig.JctHeat}c),  we measure a heating per junction traverse of $113\pm4$, in good agreement with the average heating predicted by the Doppler recooling method.  Scanning the waveform update frequency and RF amplitude, we observe a minimum dynamic heating of  $73\pm2$ quanta per trip at an update rate of $700$ kHz ~(Fig.~\ref{fig.JctHeat}d) and large on-resonance dynamic heating exceeding $300$ quanta per trip. 

Though the sideband envelope method generates repeatable and reliable results, it is vulnerable to laser and magnetic field drift as only a single point is sampled per sideband. The reliability of this method can be increased at the cost of longer experiment times by sampling more points around each sideband. This method can also be extended to models that incorporate mixtures of coherent and thermal states. Using the model of Ref.~\cite{WaltherPRL2012}, we find results that are consistent with thermal states  with a coherent contribution of less than 1 quanta.

\section{Conclusion}
\label{sec:conclusion}

We performed basic ion motion operations with $^{40}$Ca$^+$ ions in a Y-junction surface-electrode ion trap. These operations were performed reliably in this trap over two years, albeit with a noticeable degradation in performance at the loading slot. Stationary heating rates are found to be comparable with other room temperature traps without surface cleaning. We measured minimal dynamic heating for adiabatic linear transport and tens of quanta of heating through the junction. Without improved stationary heating rates or junction heating rates, these types of traps will require sympathetic cooling ions to achieve high quality two-qubit gates. This necessitates further study of the ion species choice and crystal configurations, which are critical for effective sympathetic cooling. A complementary approach would be to implement a controlled diabatic transport to reduce any coherent excitation after the junction traverse, as has been demonstrated for linear transport \cite{WaltherPRL2012, Bowler2012}.   

Although the measured heating rates through the junction are much higher than values measured in Ref.~\cite{Blakestad2009} for a two-layer X-junction, our results show that reasonable results can be obtained with commercial electronics and surface-electrode ion traps fabricated by a foundry process. With improvements to junction design and in-situ trap cleaning, it seems reasonable that mass-produced surface electrode ion traps could achieve junction heating as low as a single quanta per trip.

\section*{Acknowledgments} The authors thank D L Moehring for useful discussions. This work was supported by the MUSIQC project as part of the IARPA MQCO program under ARO contract W911NF-10-1-0231. AB was supported by the NSF through the Quantum Information for Quantum Chemistry Center for Chemical Innovation (CHE-1037992). GDV was supported by the GTRI Shackelford Fellowship.

\providecommand{\newblock}{}

\appendix
\section{Sideband Envelope Method}
\label{Appendix}
When a ground state ion in a one-dimensional harmonic potential with $n$ quanta of motion, $\left |g,n\right\rangle$,  is resonantly driven to an excited state, $\left|e,n+m\right\rangle$,  by the $m$th order sideband, the population in the excited state at time $t$ can be expressed as 
\begin{equation}
P_{n, m} =\sin^2(\Omega_{n,m} t)
\end{equation}
with the state dependent sideband Rabi frequency $\Omega_{n,m}$ given by
\begin{eqnarray}
\Omega_{n,m}&=\Omega_{0,0}~e^{-\eta^2/2} (n_<!/n_>!)^{1/2} \eta^{|m|}L_{n_<}^{|m|}(\eta^2)  
\label{pure_equation}
\end{eqnarray} 
 where $n_<$($n_>$) is the lesser (greater) of $n$ and $n+m$, $L^{\alpha}_n$ is the generalized Laguerre polynomial, and $\eta$ is the Lamb-Dicke parameter \cite{Wineland1998}. 

For an initial mixed motional state $\rho=\left|g\right\rangle\left\langle g \right| \otimes \rho_{\rm{motion}} $, the excited population for each sideband, $P_{\rho,m}$  is determined by  
\begin{equation}
P_{\rho, m} = \sum_{n = \max(0, m)}^{\infty}P_n  \sin^2(\Omega_{n,m}t) \label{mix_equation}
\end{equation}
where $P_n$ is initial population of each motional state, $P_n=\left \langle g,n\right|\rho\left |g,n\right\rangle$.  For thermal states, $P_{n}=e^{-\hbar\omega n / k_BT}(1-e^{-\hbar\omega/ k_BT})$. By fitting the measured sidebands to Eq.~\ref{mix_equation}, we can find $T$ and then calculate $\overline{n}$. For numerical convenience, we truncate the sum to $n=1000$ and use a Boltzmann distribution of the truncated motional states to calculate $P_n$.

\end{document}